\documentstyle[12pt]{article}

\large
\newcommand{\be}{\begin{eqnarray}}
\newcommand{\ee}{\end{eqnarray}}

\begin{document}

\begin{tabbing}
\`NSF-ITP-93-123\\
\`Oct. 1993
\end{tabbing}
\vbox to  0.8in{}
\centerline{\large\bf Which chiral symmetry is restored in hot QCD?}
\vskip 2cm
\centerline{\bf Edward Shuryak}
\centerline{\it Physics Department, SUNY at Stony Brook, NY 11794}
\centerline{\it and ITP, UCSB, Santa Barbara, CA 93106-4030}
\vskip 2cm
\centerline{\bf Abstract}
  We review the current status of the problem of chiral symmetry restoration,
discussing an open question of whether at the $SU(N_f)_A$
restoration point $T=T_c$
the $U(1)_A$ chiral symmetry is or is not (approximately)
restored. New  lattice and instanton-based studies are considered,
and some new calculations are suggested
to clarify the issue. We also speculate on possible experimental manifestations
 of two possible scenarios.

\newpage
\noindent
{\bf 1.The problem}
\vskip 0.3cm

For simplicity,
we ignore all effects due to the non-zero quark masses,
and consider QCD in the chiral limit, with
$N_f$ {\it massless} quarks. In this case the QCD Lagrangian is just
a sum of two separate
terms, including  right- and left-handed quarks, which implies {\it two}
chiral symmetries: $SU(N_f)_A$ and $U(1)_A$.

 Their fate is well known to be different.
 The former one is spontaneously
broken in the QCD vacuum  but it is restored at high temperatures, above
some critical point, denoted as $T=T_c$.
 Present lattice
simulations with
dynamical quarks suggest $T_c \approx 150 MeV$.

  The $U(1)_A$ chiral symmetry is not related to Goldstone bosons, as Weinberg
has first pointed out. It is now well known this is is because
this symmetry simply does not exist at quantum level, being violated by
the 'chiral anomaly'. Its physical
mechanism is also known \cite{tHooft}, it is driven by the tunneling between
the topologically different gauge vacua, described semiclassically
by the instantons \cite{Belavin_etal}.

 It is also known, that
at high temperatures the instanton-induced amplitudes are suppressed
due to the Debye-type screening \cite{Shuryak_conf,PY},
and therefore (at some accuracy level)
we expect this symmetry to be 'practically restored'
at high T. Let us denote the point where it happens
with some reasonable accuracy  as $T_{U(1)}$.

  The main question to be discussed below is the
interrelation of the two temperatures, $T_c$ and $T_{U(1)}$. Let us
refer  as 'scenario 1' to  the case $T_c \ll T_{U(1)}$ in which the
complete $U(N_f)_A$ chiral symmetry
is restored only well inside the quark-gluon plasma domain. Another
possible case\footnote{The case $T_c >> T_{U(1)}$ does not seem to be
possible.}
$T_c \approx T_{U(1)}$ is  referred below as 'scenario 2': it
implies significant changes  in many hadronic channels
around this phase transition
point. As we will discuss below, these two scenarios
lead to quite different predictions.

   This important question is well known and it
 was already discussed in literature, but still
 we do not have a definite answer. In these comment,
 we are going to look again at
 its somewhat controversial
history, and discuss what was recently done (and also can be done)
  in theory,
numerical and 'real' experiments in order to answer it.

 Pisarski and Wilczek \cite{PW}
have considered
 this question in connection with the order of the chiral phase transition.
 They have pointed out
that in the special case $N_f=2$ the
 'scenario 1' is likely to
 lead to  {\it the second order} transition. The reason is
an effective Lagrangian describing the softest modes is
essentially the Gell-Mann-Levy sigma model, same as for the O(4) spin systems.

 However, in that paper Pisarski and Wilczek have actually argued in
 favor of the 'scenario 2'.
 Their argument was as follows:
'if instantons themselves are the primary chiral-symmetry-breaking mechanism,
then it is very difficult to imagine the unsuppressed $U(1)_A$-breaking
amplitude at $T_c$'. They have even mentioned
 that this amplitude should  be at
$T_c$ at least an order of magnitude smaller than at
T=0, although no details of this estimate were
given.

  As we will discuss below, both lines of reasoning have accumulated new
evidences, so the question under consideration
appears to be even more controversial, than it was a decade ago. We hope new
studies, especially the lattice-based ones, can finally clarify it.

\vskip 0.3cm
\noindent
{\bf 2.Is the $N_f=2$ phase transition similar to that for O(4) spin system? }
\vskip 0.3cm

 During the last 2 years the large-scale
 numerical  lattice simulations with dynamical fermions
have addressed the issue of QCD phase transition, especially its
 order. It was indeed found that the case of 'many flavors'
$N_f>2$ (especially well studied case is 4) leads to sharp
first order transition, with two phase coexistence etc.,
 while   in the particular case
$N_f=2$  such phenomena were  not observed
 (see discussion and original references
in \cite{Karsch} or proceedings of yearly lattice conferences).

  Rajagopal and Wilczek \cite{RW} have  argued that
in this case one probably has a second order
transition,  analogous to that of the O(4) spin system. They have
therefore proposed a suitable effective Lagrangian, essentially
a 3-dimensional $\sigma$ model,
and  have applied it to dynamical simulations of  the phase transition.
  Now, do the available lattice data really support
 these ideas?

The most straightforward way to test them is to compare
the critical behaviour in both cases, testing whether
the $N_f=2$
QCD and the O(4) spin system do or do not belong to
the same universality class. (Similar comparison
was done in the past for many different phase transition, including
'deconfinement' one in $SU(2)_c$ pure gauge theory.)

  The first critical index to compare is the one for the order parameter, for
which
the analogy \cite{RW} suggests
\be <\bar \psi \psi > \sim | (T-T_c)/T_c|^{.38\pm.01}
\ee
Unfortunately, it is not that simple to
 compare it to lattice  data, because chiral order parameter
is only obtained as a small-mass limit, and the masses used on the lattice
 cannot be made sufficiently small. Recent analysis \cite{Karsch}
has concluded, that the data are consistent with O(4) critical exponents,
 although say O(2) ones are not also excluded.

  The second obvious question is the behaviour of global thermodynamical
quantities, such as the
 specific heat. The O(4) spin system is believed to have a very
amusing behaviour, with {\it positive} power\footnote{As far as I know, it
remains unknown whether the coefficient is positive or negative: thus one can
have a dip or a peak.}
\be C(T) \sim | (T-T_c)/T_c|^{.19\pm.06}
\ee
It means that the singular contribution of the soft modes
 {\it vanishes} at the critical point, and in order to single it out
 the 3-ed derivative
of the free energy should then be calculated.

However,  lattice data for the  $N_f=2$ QCD
actually do show a {\it huge peak} in
the specific heat around $T_c$.  It certainly implies, that
many new degrees of freedom become available
(or are significantly changed) in this region.
What these degrees of freedom are, both in hadronic language and in the
quark-gluon one, remains the major open problem in the field.

  Of course, there is no logical contradiction here: apart of large (but
smooth)
peak one may eventually find a small 'kink', which is truly singular.
And still,
if one can get any hint from the available lattice data at all,
I think they certainly indicate that
point to the o the $N_f$ QCD and
the $O(4)$ spin system are very
different physical objects, even as far as
such global parameters as thermodynamical properties are considered.

\vskip 0.3cm
\noindent
{\bf 3.Are both chiral symmetries restored simultaneously? }
\vskip 0.3cm
  We have already mentioned at the end of section 1
the arguments in favor of 'scenario 2' mentioned in \cite{PW}.
Are there any new
 arguments for or against it?

   The early
  conjuncture
(see e.g.\cite{CDG} and references therein)
 relating instantons with
spontaneous breaking of the $SU(N_f)_A$
chiral symmetry has gained strong support during the last decade.
{}From a qualitative idea one has come all the way to quantitative
understanding of many details. It is now possible to
to calculate not only
the quark condensate, but also such quantities as
  the  pion or
nucleon masses, using instanton-based model.

   The latest development has resulted in
  quite detailed studies of the QCD correlation functions.
The 'instanton liquid' model
 was shown to reproduce their
 behaviour for many mesonic and baryonic channels
\cite{SV}, in agreement
with phenomenology \cite{Shuryak_cor} and lattice
calculations \cite{Negele}.
 Direct lattice studies
\cite{Hands_Teper,Chu_Huang}
have also shown direct connection between the
quark condensate (and correlation functions) and instantons.

  Thus, there is little doubt that instantons drive chiral symmetry breaking
in the QCD vacuum.
   Unfortunately, all this development corresponds to T=0, while
very little was done for the finite T case.

   Although
the  Debye screening of large-size instantons \cite{Shuryak_conf} should
strongly suppress them at high T, what exactly happens
below $T_c$ remains unknown.
  The so called 'Pisarski-Yaffe  suppression factor'
\cite{PY}
\be dn_{inst}(T)\sim dn_{inst}(T=0)exp[-\pi^2 \rho^2 T^2({2 N_c \over 3} + {
N_f \over 3})] \ee
can
be applied {\it only} in the quark-gluon plasma phase\footnote{
Whatever simple this argument is, it was not made in the original papers.
For small size instantons 't Hooft determinant is calculated correctly
even at T=0, but the temperature-dependent corrections to it are calculated
under  specific assumption
that the heat bath is nothing but
an ideal gas of quarks and gluons. Another way to see this is
to note that the Debye screening is absent below $T_c$.}.
Taking into account available
lattice data on thermodynamics and Debye screening length,
one can now argue that this relation is applicable above
$T \approx 2 T_c$. We also know now that
the typical instanton radius of
 the 'instanton liquid' is $\rho \sim 1/3 fm$ \cite{Shuryak_82}.
 Combining the two
one  finds  instanton suppression by at least
2 orders of magnitude at $T \approx 2 T_c$.

  Now, is it possible that strong
(e.g.
by one order of magnitude, as in the
original Pisarski-Wilczek estimates) {\it instanton suppression
is  already present at $T=T_c$}?
  There are several reasons to think it is not the case. Unfortunately, none
of them is direct and sufficiently strong at the moment. Let me mention two
of them.

  The first are measurements of {\it T-dependence
of the topological susceptibility} $\chi$
in pure gauge theories. Let me remind, that
in absence of fermions, it is more or less direct
measure of the instanton density. What is found \cite{chi} demonstrate that
$\chi(T)$ remains constant (or even somewhat increases) through
 the deconfinement
phase transition, while the high-T data are so far inconclusive.
Note however, that deconfinement in pure gauge theory corresponds to relatively
high temperatures, about 240 MeV, compared to chiral restoration we discuss
in the rest of the paper.

 The second argument is the following one.
The pressure of quark-gluon plasma can be written as
$p=p^{quarks,gluons}(T)-B(T)$, where the first term is perturbative black-body
type contribution, and
the second
'bag term' describes the {\it difference} between the
 non-perturbative energy at
zero and non-zero T.
 Strong suppression of instantons at $T_c$ implies that the second
term is mostly the vacuum energy
\be B={b/128\pi^2 }<(gG^a_{\mu\nu})^2> \approx 400 MeV/fm^3
\ee
(where b is the famous coefficient of the Gell-Mann-Low beta function).
Comparing it with the perturbative term, one then finds that it is too small to
get  {\it positive pressure}
 of the plasma phase. More details on lattice data and
'melting' of quark condensate can be found in ref. \cite{Koch_Brown}, where it
is concluded that instanton suppression factor at $T_c$ is about 1/2 or so.
Although these arguments are not of course strict, they do indicate that
strong instanton suppression at $T_c$ is very unlikely.

    Moreover, the qualitative picture of
instanton-driven chiral symmetry restoration has
significantly changed since the days of original conjectures. In fact,
 {\it suppression} of instantons is not the only way
to 'kill' the quark condensate. Not only the {\it number}
of instantons is important, but also their {\it relative positions and
orientations}. If the tunneling events,
 instantons and anti-instantons, are rearranged into some finite clusters with
{\it zero topological charge}, such as well-formed
'instanton-anti-instanton molecules',
all properties of the vacuum are completely changed.
In particular, such ensemble leads to zero quark condensate, or
restored chiral symmetry.

 Numerical simulations of
interacting 'instanton liquid' at finite T \cite{SV_T}
shows that it is exactly this type of correlations which
are rapidly build up in the critical region. 'Pairing' of instantons allow them
in this case to interact all the way around the Matsubara torus.
This is especially enhanced in case of many light quark flavors, since
light quark do not propagate well into space direction at high temperatures.

   To work out a detailed theory of such correlations will take some time,
so a
schematic model for this phenomenon was recently proposed and studied
in ref. \cite{SSV_mix}. The instanton ensemble is assumed to be a {\it mixture}
of two components; (i) a random 'liquid' and (ii) $I\bar I$ molecules
\footnote{Note the difference with earlier treatment of the problem,
e.g in ref. \cite{IS}: now
 these two components are not treated as two phases,
for $T<T_c$ they are present simulteneously .}.
It is
easy to generate such ensemble, for various values of the
'molecule fraction' $f_m={2 N_{molecules} \over N_{all}}$,
 the main parameter of the model.
Behaviour of the main order parameter, $<\bar q q>$ on $f_m$ is very similar as
its T-dependence, measured on the lattice.

It was further found, that different correlation functions depend
on it quite differently.
Some correlators are rather insensitive to it.
For example, the vector one does not change by more than
about 10 \% for all compositions, from $f_m=0-1$.  Even the pion
correlator shows
remarkable stability for $f_m=0-0.8$, with subsequent rapid drop if its
coupling constant toward $f_m=1$, at which point it coinsides with
its scalar 'relatives' (see below).
At the same time,  some correlators show dramatic
sensitivity to $f_m$, especially the scalar ones.

  There is no place here for further details, let me now point out
the main physics of the phenomenon. Its one side was repeatedly emphasized
before ( see e.g. \cite{Brown_Rho}):
 as the quark condensate 'melts down', so does
a 'constituent quark mass'  (whatever it means) and hadrons
in average should become lighter. However, there is
another side of the story: the instanton-induced interaction between
quarks 'melts' as well. In the channels where this interaction is {\it
attractive}, such as $\rho,\pi$, the two effects work in the opposite
directions and can cancel each other. However, in the channels where it
is {\it repulsive}, such as axial and I=1 scalar ones, they work together
and lead to much stronger effects.

\vskip 0.3cm
\noindent
{\bf 4.What lattice observables can further clarify the problem? }
\vskip 0.3cm

    Let me start with the most obvious ones.
 For example, why not measure  the T-dependence
 of the instanton density, either 'globally', by
the topological susceptibility, or
'locally, by
'cooling' or by the fermion method?

  We have already mentioned some results on
T-dependence of the topological susceptibility in
{\it pure gauge} theory: those are not yet very accurate, and
can be much improved. The same question in theories with quark was studied
by Bernard et al \cite{sigma}, but in this
case $\chi(T)$ is simply proportional to the quark
condensate, as Word identities imply.

 It is in principle possible to see 'molecular'
contribution, substituting 4-volume by its surface in the definition of the
topological susceptibility
\be \chi_{surface}=lim_{surface\rightarrow\infty} <Q^2>
\ee

  Unfortunately, other
methods are not very sensitive to 'molecules'. 'Cooling'
can lead to 'annihilation' of a correlated pair. 'Fermion' method can help
only if the pair is not too close, so that they still lead to pretty small
eigenvalues of the Dirac operator.

Another question, which was often asked: why lattice
people do not measure directly
the $\eta'$ mass, which is obviously the one most intimately
related with the $U(1)_A$ chiral symmetry and its (practical) restoration.

The answer\footnote{It seems a strong argument, forbidding measurements of the
$\eta'$ mass: as for the correlator at not-so-large distances, I think
the needed statistical accuracy may still be achieved.}
 is technical: in such case it is clear that one should
calculate  not only the so called 'one-loop diagram'
\be K_1(x,y)= -<Tr[\Gamma S(x,y) \Gamma S(y,x)]>\ee
 but that of the 'two-loop' one
\be K_2(x,y)= <Tr[\Gamma S(x,x)]Tr [\Gamma S(y,y)]>)
\ee
 as well, and the latter is much
more difficult to evaluate because one has to invert the Dirac matrix many
times.

   However, one can come around this problem, by looking into the {\it scalar}
correlators.
  Let us discuss for simplicity only two light flavors and
use the old-fashioned notations, calling the isoscalar
I=0 scalar channel a $\sigma$ one,
and isovector $I=1$ scalar channel a $\delta$ one
\footnote{Now particle data table denote notations $f_0$ and $a_0$ to I=0,1
scalars: however particular resonances listed there under these names hardly
have anything to do with correlators under consideration.}.
Under $SU(2)_A$ transformations, $\sigma$ is mixed with $\pi$, thus restoration
of this symmetry at $T_c$ require identical correlators for these two channels.

Another chiral multiplet
is $\delta, \eta_{non-strange}$, where the last channel is the
SU(2) version of $\eta'$: at T=0 those are very heavy and are not considered
in chiral Lagrangians, or course.

   On the contrary, the
$U(1)_A$ transformations mix e.g. $\pi,\delta$ type states,
and thus its 'practical restoration' should imply that  such type of
correlators should become similar. Finally, if {\it both} chiral symmetries are
restored, a simpler statement follows: left-handed quarks never become
right-handed, therefore all $\pi,\eta_{non-strange},\sigma,
\delta$ correlators should become the same.

    Now it is clear
what one should do:
 {\it to measure the $\delta$ correlator},
 for which one does not need the double-loop
diagram \footnote{
 For its charged component (e.g. $\bar u d$) it is trivial, because
two quarks have different flavor:
 a one-line algebra shows why it is so for neutral component as well.},
as a function of T. Comparison with
the pion correlator will tell us where they become close enough, which
defines the $T_{U(1)}$
under consideration.

  Unfortunately, strong
 confusion  was going on in lattice measurements
of scalar correlators.  A number of
such measurements was
reported  \cite{sigma}, with Kogut-Susskind fermions. In this formulation
different flavors live on different lattice sites, so the
operator is quite different
for $\sigma,\delta$. Although  $\sigma$ operator
(and $\sigma$ label) was used, only the {\it one-loop}
  part $K_1^\sigma(x,y)$ was calculated.
 It was somehow implicitly assumed, that the
two-loop one, which leads to separable contribution
\be K_2^\sigma(x,y) \rightarrow <\bar q q(x)><\bar q q(y)> \ee
at large distances $|x-y|$, is something like a constant, which should be
subtracted.

   However, there is no ground for such an assumption, and experience with
$\sigma$ correlator calculated in the 'instanton liquid model' \cite{SV} shows
that. even after subtraction, there remains a significant
contribution of some light states, presumably
related to the famous sigma peak in two pion scattering. As a result,
$\sigma$ correlator look quite different from what was found in these works.

   Now come better news. The non-locality and difference between
$\sigma$ and $\delta$ operators are, after all, just lattice artifacts.
In continuum limit,  both one-loop diagrams
become identical
\be  K_1^\sigma(x,y) \approx K_1^\delta(x,y)
\ee
At T=0 we know that the
$\delta$ channel does not have any prominent resonances and
is dominated by multi-pion states with energy above 1 GeV.
And indeed, such features are seen in
 $\delta$ correlator, calculated with Wilson fermions
(e.g. \cite{Negele}).
Thus, we are able to test our conjuncture, find qualitative agreement.
 On this ground,
let me  speculate that it is also true at non-zero T,
so results of \cite{sigma}
do {\it resemble $\delta$ rather than $\sigma$ correlator}. If so, instead of
the well anticipated
 $\pi-\sigma$ degeneracy at $T_c$ they gave actually hint a toward more
interesting statement, the
$\pi-\delta$ degeneracy, which implies restoration of
 {\it both} chiral symmetries
at $T_c$ ('scenario 2').

  Of course, further studies are needed to clarify this conjuncture.
In doing this, it would be especially useful to repeat what was done
by Gottlieb et al \cite{sigma},
namely measure separately the {\it difference} between scalar and pseudoscalar
correlators, which has the most dramatic temperature (and quark mass)
dependence.

   Finally, another way of answering the question we discuss is to study
{\it T-dependence} of the average
values of certain four-fermion operators.
Selecting those which are related to chiral symmetry breaking,
one may get rid of perturbative contribution. However, the small mass
extrapolation is still necessary, as for the quark condensate.

   A quantity sensitive to $U(1)_a$ is well known, it is e.g. 't Hooft-type
operator
\be  O^{tHooft}(T) = <<[(\bar u_R u_L) (\bar d_R d_L)]>> \ee
(where L,R stand for left and right components of the quark fields).
Around or above $T_c$ there is no quark condensate, and 'unpaired'
instantons have very small
density
$O(m^{N_f})$ \cite{tHooft}. However,
in the measurement of quantities like this one, the denominators of
the quark propagators will take  powers of quark masses back, recovering
the famous 't Hooft effective interaction.
In a sense, the operator considered can induce a tunneling event by itself,
which was not present in the vacuum without it.
 Unfortunately,
 it implies in practice that small fraction of configurations will
produce a large signal: certainly not an easy way for measurements.
However, this is what one should do, trying to understand the question
under consideration.

  Let us for completeness also mention some
'alternative $SU(N_f)$ order parameters', the expectation values of
the following 4-fermion operators
\be O_1(T)= <<[\bar u_L \gamma_0 u_L-\bar d_L \gamma_0 d_L][L\rightarrow R]>>
\ee
\be O_2(T)= <<[\bar u_L \gamma_i u_L-\bar d_L \gamma_i d_L][L\rightarrow R]>>
\ee
\be O_3(T)= <<[\bar u_L \gamma_0 t^a u_L-\bar d_L \gamma_0 t^a
d_L][L\rightarrow R]>>
\ee
\be O_4(T)= <<[\bar u_L \gamma_i t^a u_L-\bar d_L \gamma_i t^a
d_L][L\rightarrow R]>>
\ee
As it is discussed in \cite{Kapusta_Shuryak}, those enter Weinberg-type
sum rules at non-zero temperatures, related with the difference between
vector and axial correlators.

\vskip 0.3cm
\noindent
{\bf 5.Can one observe  the difference
between the two scenarios in 'real' experiments?}
\vskip 0.3cm

   It is expected that in high energy heavy-ion collisions at RHIC and LHC,
the system will spend a significant amount of time, of
the order of 30 fm/c., in the so called 'mixed phase' at $T\approx T_c$.
It is not easy to produce quark-gluon plasma, but, after it is produced,
it also difficult to transform it into ordinary low-density matter.

 A very fascinating
possibility, suggested by Bjorken \cite{BJ}
and recently studied in \cite{RW} using 'quenched' scenario, is
possible formation
of large-amplitude classical pion field
due to large critical fluctuations. In order to described it
  one should know which type of fields should
be included in the effective Lagrangian, describing dynamics of the phase
transition. Scenario 1 implies that it is a (3-d) variant of sigma model, while
scenario 2 demand the inclusion of (at least) twice more soft modes.

  I could not figure out any observable signals for both scalars, $\sigma$ and
$\delta$ even if fluctuations in their directions at the transition is equally
strong to that of the pion. However,
it is clear that scenario 2 implies {\it large fluctuation in
$\eta_{non-strange}$ direction}, and those can  in principle be observed.

  First of all, as proven by the WA80 experiment,
both $\pi^0,\eta$ can be extracted in the $\gamma\gamma$ mode,
even in large-multiplicity heavy-ion collisions with its huge combinatorial
background \footnote{ Unfortunately, it can only be done statistically, so
'fluctuations
 in eta direction' on event-per-event basis can only be done indirectly,
via fluctuations in total number of photons.}.

  Moreover, the
$\eta$ spectra should have clear signs of the $U(1)_A$ restoration,
provided scenario 2 is the case. In this case $\eta$ mesons are easily
produced inside the fireball, being very light, but then,
while trying
to leave the system, they
should then experience
the influence of very strong collective potential.
Its crude estimate is  $V\sim (m_\eta-m_\pi) \sim 400 MeV $.
 In contrast to pions,
for which it is much smaller, it is comparable to their kinetic energy at
breakup $E_{kinetic}\approx 3T\sim 400 MeV$.
 As a result, a significant fraction of etas should be trapped inside the
fireball, and
eventually shifted to much lower kinetic energies. This phenomenon,
analogous to electrons evaporated from the hot cathode, should lead to
a peak in the eta spectrum at small $p_t$.
\vskip .3cm
\noindent
{\bf 6.Summary}
\vskip 0.3cm
   Summarizing our discussion, we can repeat that the main question discussed
in these comments remain open.

   At one hand, it looks plausible that 'phase transition in mathematical
sense', the singularity, is related with 4 'soft' modes, pions and sigma
ones.
  At another, it becomes increasingly clear that physics at $T\approx T_c$
cannot be described in terms of these fields alone, and excitations of matter
in this region are very different from those at $T=0$.

  In particular, there are evidences that {\it parity partners} of pions and
 sigma, $\delta, \eta_{non-strange}$,
which are very heavy or even non-existing at T=0, become comparably
light. Such approximate $U(1)_a$ restoration is not however
based
 on 'instanton suppression', but rather on new mechanism,
 a reorganization of the 'liquid' into into the
instanton-anti-instanton molecules.

  Finally, if these phenomena take place, large fluctuation
of $\eta$ production in heavy ion collision,
especially at small transverse momenta, are predicted.

\vskip .3cm
\noindent
{\bf 7.Acknowledgements}
\vskip 0.3cm
 Lattice aspects of the problem were explained to me
by F.Karsch,B.Petersson,R.Sugar and D.Toissaint.
  Discussion of relevant  questions took place
in ITP, Santa Barbara,  during the
 'Finite temperature QCD' program. The hospitality in ITP and
financial support by NSF are greatly acknowledged.

% this is bib.tex

\end{document}